\documentclass[aps,preprint,epsfig,rotate]{revtex4}
\usepackage{graphicx}
\usepackage{bm}
\usepackage{epsfig}

\input epsf
\begin{document}
\title{On the hyperfine structure of the triplet $n^{3}S-$states of the four-electron atoms and ions}

\author{Alexei M. Frolov}
\email[E--mail address: ]{afrolov@uwo.ca}

\affiliation{ITAMP, Harvard-Smithonian Center for Astrophysics, \\
         MS 14, 60 Garden Street, Cambridge MA 02138-1516, USA}  

\affiliation{Department of Applied Mathematics \\
       University of Western Ontario, London, Ontario N6H 5B7, Canada} 

\date{\today}

\begin{abstract}

Hyperfine structures of the triplet $n^3S-$states in the four-electron Be-atom(s) and Be-like ions are considered. It is shown that to determine the hyperfine structure 
splitting in such atomic systems one needs to know the triplet electron density at the central atomic nucleus $\rho_T(0)$. We have developed the procedure which allows 
allows one to determine such an electron density $\rho_T(0)$ for arbitrary four-electron atoms and ions.   

\noindent 
PACS number(s): 31.30.Gs, 31.15.vj and 32.15.Fn

\end{abstract}

\maketitle
\newpage

\section{Introduction}

In this communication we develop the new ab-initio method which can be applied for accurate evaluation of the hyperfine structure splittings in the triplet ${}^3S-$states 
of the four-electron atoms and/or ions. As follows from experiments such triplet $S(L = 0)-$states have an interesting hyperfine structure. For simplicity, let us consider, 
the triplet $2^3S-$state of the four-electron beryllium atom(s) (${}^{7}$Be, ${}^{9}$Be and ${}^{\infty}$Be). In general, if $F$ is the total electron-nuclear spin of the 
$2^3S-$state of an atom and $I_N$ is the spin of atomic nucleus (Be) and $I_N \ge 1$, then in experiments one can observe splitting of this state into a triplet of states. 
The total spin of these states equals $F = I_N + 1, I_N$ and $I_N - 1$, respectively. If $I_N = \frac12$, then we can see only a doublet of states with $F = \frac12$ and 
$\frac32$. This method can also be used to determine the hyperfine structure splitting for an arbitrary bound (triplet) $n^3S-$states in four-electron atoms and ions. This 
includes the triplet $2^3S-$state of the four-electron Be atoms (different isotopes). Our analyisis of the hyperfine structure of the four-electron atoms and ions is based 
on the generalization of the method developed earlier by Fermi \cite{Fermi} for the doublet $n^{2}S-$states of three-electron Li-atom(s) and Li-like ions (see also 
\cite{Fro2016}).

First, we need to introduce the triplet electron density in a few-electron atom/ion. Formally, the triplet electron density is the spatial two-electron density distribution 
of the two atomic electrons which form one triplet pair. If we have a number of such pairs in atom/ion, then we need take into account all triplet electron pairs. Singlet 
electron pairs do not contribute to the triplet electron density. In reality, for accurate evaluations of the hyperfine structure splitting in the triplet states of 
four-electron atoms/ions one needs to know the triplet electron density at the central atomic nucleus which has non-zero electric charge $Q e$. The general definition of the 
electron density at the central atomic nucleus is written in from $\rho(0) = \sum^{N_e}_{i=1} \langle \delta({\bf r}_{iA}) \rangle$, where $i$ is the electron's index, while 
index $A$ means the central atomic nucleus. For instance, for the singlet ground $1^1S-$state in the two-electron helium atom one finds $\rho_{S}(0) \approx$ 
1.8104293185013928, while for the triplet $2^3S-$state of the helium-3 atom we have $\rho_{T}(0) \approx$ 1.31963500836957 \cite{Fro2007}. The last numerical value leads to 
the following hyperfine structure splitting in the $2^3S-$state of the ${}^{3}$He atom: $\Delta E_{hss}$ = 6740.452154 $MHz$ \cite{Fro2007} (see also \cite{BS}). The 
corresponding experimental value is $\Delta E_{hss}$ = 6739.701171(16) $MHz$ \cite{PRA70}.  

However, such a definition of the electron density cannot be used for the triplet states in atoms/ions, if the total number of bound electrons exceeds two. The reason is obvious, 
since all atoms/ions with more than two bound electrons always have the shell electronic structure. This means that the internal electrons form a number of closed electron shells 
which have zero spin, i.e. singlet electron shells. The electrons from the outer-most shell(s) can interact with the nuclear spin $I_N$, its numerical value it differs from zero. 
In general, this leads to the appearance of the hyperfine structure splitting in atoms and ions, if the total spin of outer-most electrons exceeds zero. This leads to the 
appearance of the hyperfine structure splitting in $N_e-$electron atoms and ions, where $N_e \ge 3$. It is clear that in actual atoms and ions we have small interactions between 
electrons from internal and outermost electron shells, e.g., electronic correlations, spin-spin interactions, etc. As follows from this picture the analysis and numerical 
computations of the hyperfine structure splitting are significantly more complicated than for the two-electron helium atom. In the lowest-order approximation we need to define 
the triplet electron density at the atomic nucleus must be given in a different manner. An alternative definition of the triplet electron density at the atomic nucleus ($A$) can 
be written in the form (see, e.g., \cite{Fermi}, \cite{Fro2016}, \cite{Lars})        
\begin{equation}
  \rho_T(0) = \sum^{N_e}_{i=1} \langle \delta({\bf r}_{iA}) (\sigma_{z})_i \rangle = \langle \Psi \mid \sum^{N_e}_{i=1} \delta({\bf r}_{iA}) (\sigma_{z})_i 
  \mid \Psi \rangle  \; \; \; \label{eq1}
\end{equation}
where $\delta({\bf r}_{iA})$ is the electron-nucleus delta-function (the symbol $A$ designates the atomic nucleus) and $(\sigma_{z})_i$ is the $\sigma_z$ matrix of the $i$-th 
atomic electron, i.e. $(\sigma_{z})_i \alpha(i) = \alpha(i)$ and $(\sigma_{z})_i \beta(i) = - \beta(i)$ (see, e.g, \cite{LLQ}, \cite{Dirac}). In Eq.(\ref{eq1}) and everywhere below 
we assume that the wave function of the bound $2^{3}S-$state of the four-electron atom/ion has unit norm. As follows from Eq.(\ref{eq1}) the triplet electron density $\rho_T(0)$ 
equals zero identically for an arbitrary singlet state in a few-electron atom, including the two-electron He atom. For the triplet states in the helium-3 atom the new definition of 
the electron density, Eq.(\ref{eq1}), leads to the same hyperfine structure splitting as mentioned above. For the ground $2^{2}S-$state in the three-electron Li atoms and analogous 
ions such a definition of the doublet electron density at the central atomic nucleus, Eq.(\ref{eq1}), allows one to evaluate the correct numerical values of the hyperfine structure 
splittings (see, e.g., \cite{Fro2016}) which are in good agreement with the known experimental values \cite{Liatom}.     

By using this definition of the triplet electron density at the atomic nucleus, Eq.(\ref{eq1}), we can write the following formula (Fermi-Segr\'{e} formula (see, e.g., \cite{LLQ})) 
for the hyperfine structure splitting of the $2^3S-$states in the Be atom (see, e.g., \cite{LLQ})
\begin{equation}
  \Delta E_{hf} = \frac{8 \pi \alpha^2}{3} \mu_B \mu_N g_e g_N \; \; \rho_T(0) \; \; \frac12 [F(F + 1) - I_N(I_N + 1) - S_e(S_e + 1)] \; \; \label{eq2}
\end{equation} 
where ${\bf S}_{e}$ is the total electron spin of the atom, ${\bf I}_{N}$ is the spin of the nucleus in those isotopes of the Be atom(s) for which $\mid {\bf I}_{N} \mid \ne 0$ and
${\bf F}$ is the total angular momentum operator ${\bf F} = {\bf L} + {\bf S} = {\bf S}_e + {\bf I}_N$ of the four-electron atom/ion. For the triplet $S-$states in the four-electron 
atoms/ions the vector-operator ${\bf F} = {\bf S}_e + {\bf I}_N$ can be considered as the total spin of the atom, i.e. the sum of the electron and nuclear spins. Also, in this 
formula $\alpha \approx 7.297352568 \cdot 10^{-3} (\approx \frac{1}{137})$ is the dimensionless fine structure constant, $\mu_B$ is the Bohr magneton ($\mu_B = \frac12$ in atomic 
units) and $\mu_N = \mu_B \frac{m_e}{m_p}$, where $\frac{m_p}{m_e}$ = 1836.15267261 is the ratio of the proton and electron masses. The notation $g_e$ in Eq.(\ref{eq2}) means the 
electron gyromagnetic ratio $g_e$ = -2.00223193043718 \cite{CRC}. The factor $g_N$ for the ${}^9$Be nucleus is $g_{N} = \frac{f_N}{I_N} \approx$ -1.177432 $\Bigl(\frac23\Bigr)$ = 
-0.7849547, since $f_N = -1.177432$ and $I_N = \frac32$. Finally, the formula for the hyperfine structure splitting $\Delta E_{hf}$ in the $2^3S-$state of the four-electron ${}^{7}$Be 
and ${}^{9}$Be atoms takes the form
\begin{eqnarray}
 \Delta E_{hf}(MHz) = 314.061338965 \; \rho_T(0) \; [F(F + 1) - I_N(I_N + 1) - S_e(S_e + 1)] \label{eq3}
\end{eqnarray}
where the factor 6.579 683 920 61$\cdot 10^9$ ($MHz/a.u.$) has been used to re-calculate the $\Delta E_{hf}$ energy from atomic units to $MegaHertz$. The Fermi-Segr\'{e} formula, 
Eq.(\ref{eq2}) and Eq.(\ref{eq3}), is correct for all $S-$bound triplet states in the four-electron Be-atom(s) and Be-like ions. The same formula is applied to other four-electron 
atoms and ions which can be found in the bound triplet $S-$states. As follows from Eq.(\ref{eq3}) in order to determine the numerical value of the hyperfine structure splitting 
$\Delta E_{hf}(MHz)$ one needs to evaluate the electron triplet density at the atomic nucleus $\rho_T(0)$. Accurate numerical evaluation of the $\rho_T(0)$ value is the main goal of 
this study. To perform such an evaluation we need to construct the accurate wave functions of the triplet $S-$states of the four-electron atoms and ions. In general, these wave 
functions are obtained as the solutions of the corresponding Schr\"{o}dinger equation for the bound atomic state(s), i.e. $H \Psi = E \Psi$, where $H$ is the Hamiltonian operator 
defined below (see Eq.(\ref{Hamil})) and $\Psi$ is the wave function and $E (< 0)$ is the total energy of the bound atomic state. This problem is considered in detail in the next 
Section.

\section{Hamiltonian and bound state wave functions} 

In the lowest-order approximation upon the fine-structure constant $\alpha$ we can consider the non-relativistic Schr\"{o}dinger equation. The non-relativistic Hamiltonian $H$ of an 
arbitrary four-electron atomic system (i.e. atom, or ion) is written in the form \cite{LLQ}
\begin{eqnarray}
   H = -\frac{\hbar^2}{2 m_e} \Bigl[ \sum^{4}_{i=1} \nabla^2_i + \frac{1}{M_A} \nabla^{2}_{5} \Bigr] - \sum^4_{i=1} \frac{Q e^2}{r_{i5}} +  \sum^{3}_{i=1} \sum^{4}_{j=2 (j>i)} 
  \frac{e^2}{r_{ij}} \label{Hamil}
\end{eqnarray}
where $\hbar$ is the reduced Planck constant, $m_e$ is the electron mass, $e$ is the absolute value of the electric charge of the electron. Also, in this equation $Q e$ and $M_A$ are the 
electric charge and mass of the nucleus ($M_n \gg 1$) expressed in $e$ and $m_e$, respectively. Below, we consider the beryllium atom with the infinitely heavy atomic nucleus, i.e. when
$M_A = \infty$, or $\frac{1}{M_A} = 0$ in Eq.(\ref{Hamil}). We also discuss a few isotopes of the Be-atom with the finite nuclear masses $M_A$. In Eq.(\ref{Hamil}) and everywhere below 
in this study the subscript 5 denotes the atomic nucleus, while subscripts 1, 2, 3 and 4 stand for electrons. Note that the four-electron Be atom has two independent series of bound 
states: singlet states and triplet states. The multiplicities of these states equal $2 \cdot 0 + 1 = 1$ (singlet) and $2 \cdot 1 + 1 = 3$ (triplet). Below, we consider only the triplet 
bound states in the Be atom(s) and Be-like ions.  

To determine the bound state wave function of the Be atom in its $2^{3}S-$state we need to solve the corresponding Schr\"{o}dinger equation for the bound state(s): $H \Psi = E \Psi$, 
where $H$ is the Hamiltonian operator from Eq.(\ref{Hamil}), while $E (< 0)$ is the total energy of the $2^{3}S-$state in the Be-atom. It is clear that the numerical value of $E$ must 
be lower than the total energy of the ground $2^2S-$state of the three-electron Be$^{+}$ ion $E \approx$ -14.3247631764657 $a.u.$ (otherwise, the $2^{3}S-$state in the Be-atom will be 
unstable, i.e. unbound). Now, consider the explicit construction of the trial wave function $\Psi$. In general, the wave functions of the bound $n^3S$-states in the Be atom are
represented as the sum of products of the radial and spin functions. Each of these radial and/or spin functions depends upon spatial and spin coordinates of all four electrons. For the 
triplet states we can use only spin functions with $S = 1$ and $S_z = 1$, where $S$ is the total electron spin, i.e. ${\bf S} = {\bf s}_1 + {\bf s}_2 + {\bf s}_3 + {\bf s}_4$, of 
four-electrons and $S_z$ is its $z-$projection. Therefore, our spin function $\chi_{11}(1,2,3,4)$ is defined by the following equalities: ${\bf S}^2 \chi_{11}(1,2,3,4) = 1 (1 + 1) 
\chi_{11}(1,2,3,4) = 2 \chi_{11}(1,2,3,4)$ and $S_z \chi_{11}(1,2,3,4) = \chi_{11}(1,2,3,4)$. In general, there are two spin functions for each four-electron atom/ion in the triplet 
state. Below, we chose such functions in the form $\chi^{(1)}_{11} = \alpha \beta \alpha \alpha - \beta \alpha \alpha \alpha$ and $\chi^{(2)}_{11} = 2 \alpha \alpha \beta \alpha 
- \beta \alpha \alpha \alpha - \alpha \beta \alpha \alpha$. 

Finally, the total four-electron wave function of the triplet states in four-electron atoms and ions is represented in the form  
\begin{eqnarray}
 \Psi = {\cal A}_e [\psi(A;\{r_{ij}\}) (\alpha \beta \alpha \alpha - \beta \alpha \alpha \alpha)] + {\cal A}_e [\phi(B;\{r_{ij}\}) (2 \alpha 
 \alpha \beta \alpha - \beta \alpha \alpha \alpha - \alpha \beta \alpha \alpha)] \label{equat2}
\end{eqnarray}
where the notation $\{r_{ij}\}$ designates the ten relative coordinates (electron-nuclear and electron-electron coordinates) in the four-electron Be atom, while the notation ${\cal A}_e$ means 
the complete four-electron antisymmetrizer. The explicit formula for the ${\cal A}_e$ operator is 
\begin{eqnarray}
 {\cal A}_e = \hat{e} - \hat{P}_{12} - \hat{P}_{13} - \hat{P}_{23} - \hat{P}_{14} - \hat{P}_{24} - \hat{P}_{34} + \hat{P}_{123} + \hat{P}_{132} + \hat{P}_{124} 
 + \hat{P}_{142} + \hat{P}_{134} + \hat{P}_{143} \nonumber \\ 
 + \hat{P}_{234} + \hat{P}_{243} - \hat{P}_{1234} - \hat{P}_{1243} - \hat{P}_{1324} - \hat{P}_{1342} - \hat{P}_{1423} 
 - \hat{P}_{1432} + \hat{P}_{12} \hat{P}_{34} + \hat{P}_{13} \hat{P}_{24} + \hat{P}_{14} \hat{P}_{23} \label{equat3}
\end{eqnarray}
Here $\hat{e}$ is the identity permutation, while $\hat{P}_{ij}$ is the permutation of the spin and spatial coordinates of the $i-$th and $j-$th identical particles. Analogously, the notations 
$\hat{P}_{ijk}$ and $\hat{P}_{ijkl}$ stand for the consequtive permutations of the spin and spatial coordinates of the three and four identical particles (electrons). In real calculations one 
needs to know the explicit expressions for the spatial projectors only. 

These spatial projectors can be obtained, e.g., by applying the ${\cal A}_e$ operator to each component of the wave function in Eq.(\ref{equat2}). At the second step we need to determine the 
scalar product (or spin integral) of the result and incident spin function. After the integration over all spin variables one finds the corresponding spatial projector. For instance, in the 
case of the first term in Eq.(\ref{equat2}) we obtain the following spatial projector for the $\psi-$components of the total wave function
\begin{eqnarray}
 {\cal P}_{\psi\psi} = \frac{1}{2 \sqrt{6}} (2 \hat{e} + 2 \hat{P}_{12} - \hat{P}_{13} - \hat{P}_{23} - \hat{P}_{14} - \hat{P}_{24} - 2 \hat{P}_{34} - 2 \hat{P}_{12} \hat{P}_{34}
 - \hat{P}_{123} - \hat{P}_{124} - \hat{P}_{132} \nonumber \\
 - \hat{P}_{142} + \hat{P}_{134} + \hat{P}_{143} + \hat{P}_{234} + \hat{P}_{243} + \hat{P}_{1234} + \hat{P}_{1243} + \hat{P}_{1342} + \hat{P}_{1432}) \label{equat4}
\end{eqnarray}
Analogous formulas have been found \cite{FroWa2010} for two other spatial projectors ${\cal P}_{\psi\phi} = {\cal P}_{\phi\psi}$ and ${\cal P}_{\phi\phi}$. These formulas for the 
${\cal P}_{\psi\phi} = {\cal P}_{\phi\psi}$ and ${\cal P}_{\phi\phi}$ spatial projectors are significantly more complicated and they are not presented here (they can be found, e.g., in 
\cite{FroWa2010}). In actual bound state calculations we can always restrict ourselves to one spin function $\chi^{(1)}_{11}$ (or one spin configuration) and use the formula, Eq.(\ref{equat4}). 

The functions $\psi(A;\{r_{ij}\})$ and $\phi(B;\{r_{ij}\})$ in Eq.(\ref{eq2}) are the radial parts (or components) of the total wave function $\Psi$. For the bound states in various five-body 
systems these functions are approximated with the use of the KT-variational expansion written in ten-dimensional gaussoids \cite{KT}. Each of the spatial basis function in the KT expansion 
depends upon ten relative coordinates $r_{ij}$ only \cite{KT}. Here and everywhere below the notation $r_{ij} = \mid {\bf r}_i - {\bf r}_j \mid = r_{ji}$ means the $(ij)-$relative coordinate, 
i.e. the scalar distance between the particles $i$ and $j$ (${\bf r}_i$ are the corresponding Cartesian coordinates of the $i$-th particle). For instance, for the $\psi(A;\{r_{ij}\})$ function 
we have
\begin{eqnarray}
 \psi(A;\{r_{ij}\}) = {\cal P} \sum^{N_A}_{k=1} C_K \exp(-\sum_{ij} a_{ij} r^{2}_{ij}) \label{equat5}
\end{eqnarray}
where $N_A$ is the total number of basis function used in radial expansion, $C_k$ are the linear variational coefficients and ${\cal P} = {\cal P}_{\psi\psi}$ is the spatial projector defined 
by Eq.(\ref{equat4}). The notations $A$ (or notations $A$ and $B$ in Eq.(\ref{equat2})) stands for the corresponding set of the non-linear parameters $\{ a^{(k)}_{ij} \}$ (and 
$\{ b^{(k)}_{ij} \}$) in the radial wave functions, Eq.(\ref{equat5}) (or Eq.(\ref{eq2})). In actual calculations these two sets of non-linear parameters are optimized independently of each 
other. In general, the KT-variational expansion was found to be very effective for various few-body systems known in atomic, molecular and nuclear physics. A large number of fast algorithms 
have been developed recently for optimization of the non-linear parameters in the trial wave functions, Eq.(\ref{equat5}), allow one to approximate the total energies $E$ and variational wave 
functions $\Psi$ to high and very high accuracy. The knowledge of the highly accurate wave function can be used to determine a large number of bound state properties. For the $2^{3}S-$state
of the four-electron beryllium atom (${}^{\infty}$Be) some of the computed bound state properties expressed in atomic units can be found in Table I. The overall accuracy obtained for these 
expectation values is high, but there is a general problem related to the shell electronic structure of all few-electron atoms/ions, where the total number of bound electrons exceeds two. 
Indeed, in all current procedures the sums of all electron-nuclear and electron-electron expectation values are devided by the factors $N_e$ and $N_e (N_e + 1)/2$, respectively, where $N_e$ 
is the total number of bound electrons. Finally, all traces of the shell electronic structure in few-electron atoms/ions are lost from the results of such calculations. It is clear that 
few-body logic does not work well for actual atoms/ions with multi-shell electronic structure. This situation must be corrected in the future.   
 
\section{Hyperfine structure splitting}

Let us evaluate the hyperfine structure splitting for the triplet $2^{3}S-$state of the ${}^{9}$Be atom. In our calculations we shall apply the numerical value of the electron density 
$\rho_T(0)$ at the atomic nucleus determined with the use of the formula
\begin{eqnarray}
  \rho_T(0) = {\cal C} \langle {\cal A}_s (\Psi \chi^{(1)}_{11}) \mid \Bigl[ \sum^{4}_{i=1} \delta({\bf r}_{iA}) (\sigma_{z})_i \Bigr] \Psi \chi^{(1)}_{11} \rangle \; \; \; \label{eq7}
\end{eqnarray}
where ${\cal C}$ is a normalization constant. Numerical computations of the overlap integrals between spin-functions included in Eq.(\ref{eq7}) is significantly more complicated than for 
three-electron atomic systems. Indeed, the total number of terms in the left-hand side wave function of the Eq.(\ref{eq7}) equals 24 and each of these terms must be multiplied by four 
(number of the electron-nucleus delta-functions). This means that we have 96 terms which contribute to the numerical value of the triplet electron density at the central atomic nucleus 
$\rho_T(0)$. Formally, it is difficult to present here all details of analytical computations of the electron density $\rho_T(0)$. However, we can illustrate such computations by 
considering the two terms which can be found in Eq.(\ref{eq7}). First, consider the term in Eq.(\ref{eq7}) which contains the permutation operator $\hat{P}_{13}$ (see, Eq.(\ref{equat4}). 
Action of this operator on the spin function $\chi^{(1)}_{11} = \alpha \beta \alpha \alpha - \beta \alpha \alpha \alpha$ produces the function $\alpha \beta \alpha \alpha - \alpha \alpha 
\beta \alpha$. On the other hand, the explicit expression in the right-hand side of Eq.(\ref{eq7}) takes the form
\begin{eqnarray}
  \Bigl[ \sum^{4}_{i=1} \delta({\bf r}_{iA}) (\sigma_{z})_i \Bigr] \Psi \chi^{(1)}_{11} &=& \frac12 \delta({\bf r}_{1A}) (\alpha \beta \alpha \alpha + \beta \alpha \alpha \alpha)
  - \frac12 \delta({\bf r}_{2A}) (\alpha \beta \alpha \alpha + \beta \alpha \alpha \alpha) \nonumber \\
  &+& \frac12 \delta({\bf r}_{3A}) (\alpha \beta \alpha \alpha - \beta \alpha \alpha \alpha) 
  + \frac12 \delta({\bf r}_{4A}) (\alpha \beta \alpha \alpha - \beta \alpha \alpha \alpha) \Psi \label{eq71}
\end{eqnarray}   
where $\Psi$ is the spatal part of the total wave function. The following integration over spin variables leads to the formula for the matrix elements
\begin{eqnarray}
  \frac12 \Bigr[ \langle \hat{P}_{13} \Psi_1 \mid \delta({\bf r}_{1A}) \Psi_2 \rangle - \langle \hat{P}_{13} \Psi_1 \mid \delta({\bf r}_{2A}) \Psi_2 \rangle 
  + \langle \hat{P}_{13} \Psi_1 \mid \delta({\bf r}_{3A}) \Psi_2 \rangle + \langle \hat{P}_{13} \Psi_1 \mid \delta({\bf r}_{4A}) \Psi_2 \rangle \Bigr] \label{eq73}
\end{eqnarray}
which contains only integrals over spatial variables. Analogously, the term which include the $\hat{P}_{12} \hat{P}_{34}$ permutation operator produces the result 
\begin{eqnarray}
  \hat{P}_{12} \hat{P}_{34} \chi^{(1)}_{11}  = - \alpha \beta \alpha \alpha + \beta \alpha \alpha \alpha \label{eq75}
\end{eqnarray}
which leads to the following formula
\begin{eqnarray}
  - \langle \hat{P}_{12} \hat{P}_{34} \Psi_1 \mid \delta({\bf r}_{3A}) \Psi_2 \rangle - \langle \hat{P}_{12} \hat{P}_{34} \Psi_1 \mid \delta({\bf r}_{4A}) \Psi_2 \rangle \label{eq77}
\end{eqnarray}
The explicit integration over electron spin variables of all other terms in Eq.(\ref{eq7}) can be performed analogously. Note that the expectation values of some of the terms in Eq.(\ref{eq7})
equal zero identically, e.g., for the $\hat{P}_{1324}$ and $\hat{P}_{1342}$ permutation operators, but the final formula for the $\rho_T(0)$ value still contains many dozens of terms. It is 
clear that the same formulas for the arising spatial projectors can also be applied for various variational expansions used for accurate calculations of the four-electron Be atom, e.g., for 
the exponential variational expansion and/or for the Hylleraas expansion. In other words, the multi-dimensional gaussoids used in this study is only one possible choice of the radial basis 
wave functions. 

In our calculations of the $2^3S$-state in the  four-electron Be atom we have found the following numerical value of the triplet electron density at the atomic nucleus $\rho_T(0) \approx  
0.7404721$. With this value Eq.(\ref{eq3}) can be written in the form 
\begin{eqnarray}
 \Delta E_{hf}(MHz) &\approx&  232.553659 \; [F(F + 1) - I_N(I_N + 1) - S_e(S_e + 1)] = 232.553659 \nonumber \\
 & & \times [F(F + 1) - \frac{23}{4}] \label{eq95}
\end{eqnarray}
for the $2^3S-$state in the ${}^{9}$Be atom, where $f_N = -1.177432$ and $I_N = \frac32$. From this equation one finds that for the $2^3S-$state in the ${}^{9}$Be atom the hyperfine structure 
levels are $\varepsilon(F = \frac12)$ = -1162.768295 $MHz$, $\varepsilon(F = \frac32)$ = -465.107318 $MHz$ and $\varepsilon(F = \frac52)$ = 697.660978 $MHz$, respectively. These three levels 
with different energies determine the hyperfine structure of the ${}^{9}$Be atom. The differences between them are the corresponding hyperfine structure splittings. Analogously, for the 
${}^{7}$Be atom, where $f_N = -1.39928$ and $I_N = \frac32$ one finds the hyperfine structure levels are $\varepsilon(F = \frac12)$ = -1381.853407 $MHz$, $\varepsilon(F = \frac32)$ = -552.741363 
$MHz$ and $\varepsilon(F = \frac52)$ = 829.112044 $MHz$, respectively.

\section{Conclusion}

We have developed the new ab-initio method which can be used to determine the hyperfine structure splitting of the bound triplet bound $S-$states of four-electron atoms and ions. Our method is 
based on the explicit derivation of the analytical formula for the operator $\sum^{4}_{i=1} \delta({\bf r}_{iA}) (\sigma_{z})_i$ in the case of the four-electron spin function $\chi^{(1)}_{11} 
= \alpha \beta \alpha \alpha - \beta \alpha \alpha \alpha$. This allows us to evaluate the hyperfine structure splitting in the triplet $2^{3}S-$state of the ${}^{9}$Be and ${}^{7}$Be atoms. It 
is clear that our method used in calculations of the hyperfine structure splitting for the triplet $2^{3}S-$state of the Be atom can easily be generalized to other four-electron atoms and ions 
which have at least one stable ${}^{3}S$-state. Unfortunately, at this time the direct comparison with the experimental results for the triplet $2^{3}S-$state in the ${}^{9}$Be and ${}^{7}$Be 
atoms is not possible, since the hyperfine structure splittings for the $2^{3}S-$state in the ${}^{9}$Be and ${}^{7}$Be atoms have never been measured. On the other hand, we note that a large 
number of experiments have been performed to observe the hyperfine structure splitting of various $P-, D-$ and other rotationally excited states in the Be atoms and Be-like ions (see, e.g., 
\cite{PhysRevA1} - \cite{CPL} and references therein). Similar experiments for the triplet $2^{3}S-$state in the ${}^{9}$Be and ${}^{7}$Be atoms are urgently needed, since their results can 
help to correct an additional factor(s) used in the formula, Eq.(\ref{eq95}). Such factors can arise, since there is an obvious difference between the doublet $2^{2}S-$state of the Li atom and 
the triplet $2^{3}S-$state of the Be atom. Briefly this means that in contrast with the Li atom, in the four-electron Be-atom there are two interacting outer-most electrons and each of these 
electrons contribute to the actual hyperfine structure splitting.     

Note also that there are a few steps in our procedure which must be improved in the future computations. First, we need to improve the current accuracy of our wave functions. This means the
better numerical accuracy of the wave functions constructed from the multi-dimensional (or ten-dimensional) gaussoids. It would be nice to use other basis sets of spatial functions in such 
calculations, since this can drastically improve the overall accuracy of the whole procedure. Second, in our current computations the second spin function $\chi^{(2)}_{11} = 2 \alpha \alpha 
\beta \alpha - \beta \alpha \alpha \alpha - \alpha \beta \alpha \alpha$ is not used. Very likely, this also reduces our overall accuracy even further. Nevertheless, this study indicates clearly 
that direct computation of the hyperfine structure splitting of the bound triplet $n^{3}S-$states of the four-electron atoms and ions are possible, since the corresponding analytical expression 
for the triplet electron density at the atomic nucleus has been derived. In the future our procedure will be modified to include two (or more) spin functions.   

\section{Acknowledgments}

This work was supported in part by the NSF through a grant for the Institute for Theoretical Atomic, Molecular, and Optical Physics (ITAMP) at Harvard University and the Smithsonian Astrophysical
Observatory. Also, I wish to thank James Babb (ITAMP) and David M. Wardlaw for stimulating discussion.

\newpage


 \begin{table}[tbp]
   \caption{The expectation values of a number of electron-nuclear ($en$) and electron-electron ($ee$) properties (in $a.u.$) of the $2^{3}S-$state 
            of the neutral Be (${}^{\infty}$Be) atom.}
     \begin{center}
     \begin{tabular}{| c | c | c | c | c | c | c |}
       \hline\hline          
 atom/ion  & $\langle r^{-2}_{eN} \rangle$ & $\langle r^{-1}_{eN} \rangle$ & $\langle r_{eN} \rangle$ & $\langle r^2_{eN} \rangle$  & $\langle r^3_{eN} \rangle$ & $\langle r^4_{eN} \rangle$ \\
     \hline
 Be ($N$ = 800)    &  14.274895 & 2.0359815 & 2.6332278 & 17.19102 & 148.569 & 1472.9 \\

 Be ($N$ = 1000)   &  14.274896 & 2.0359818 & 2.6332253 & 17.19098 & 148.569 & 1472.9 \\
  
 Be ($N$ = 1200)   &  14.274897 & 2.0359820 & 2.6332235 & 17.19095 & 148.568 & 1472.8 \\
     \hline\hline     
 atom/ion  &  $\langle r^{-2}_{ee} \rangle$ & $\langle r^{-1}_{ee} \rangle$ & $\langle r_{ee} \rangle$ & $\langle r^2_{ee} \rangle$ &  $\langle r^3_{ee} \rangle$ &  $\langle r^4_{ee} \rangle$ \\
     \hline
 Be ($N$ = 800)    &  1.50343930 & 0.6192542 & 4.7138462 & 35.22448 & 320.596 & 3287.4 \\

 Be ($N$ = 1000)   &  1.50343917 & 0.6192544 & 4.7138420 & 35.22442 & 320.595 & 3287.4 \\
  
 Be ($N$ = 1200)   &  1.50343901 & 0.6192548 & 4.7138398 & 35.22435 & 320.593 & 3287.4 \\
      \hline\hline             
 atom/ion  &  $E$ & $\langle \frac12 p^2_{e} \rangle$ & $\langle \frac12 p^2_{N} \rangle$ & $\langle \delta_{eN} \rangle$ & $\langle \delta_{ee} \rangle$ &  $\langle \delta_{eee} \rangle$  \\ 
     \hline
 Be ($N$ = 800)    &  -14.430029018 & 3.60753776 & 14.89301041 & 8.740425 & 0.265321 & 0.0 \\

 Be ($N$ = 1000)   &  -14.430029235 & 3.60753825 & 14.89301164 & 8.740425 & 0.265321 & 0.0 \\
  
 Be ($N$ = 1200)   &  -14.430029456 & 3.60753933 & 14.89301235 & 8.740427 & 0.265319 & 0.0 \\
     \hline\hline
  \end{tabular}
  \end{center}
   \end{table}


\begin{thebibliography}{01}

\bibitem{Fermi}E. Fermi, Zeits. f\"{u}r Physik {\bf 60}, 320 (1930).

\bibitem{Fro2016}A.M. Frolov, JETP Lett. {\bf 103}, 739 (2016).

\bibitem{Fro2007}A.M. Frolov, J. Chem. Phys. {\bf 126}, 104302 (2007).

\bibitem{BS}H.A. Bethe and E.E. Salpeter, \textit{Quantum Mechanics of One- and Two-Electron Atoms} (Plenum Publ. Corp., New York, 1977), \S 40.

\bibitem{PRA70}S.D. Rosner and A.M. Pipkin, Phys. Rev. A {\bf 1}, 571 (1970).

\bibitem{Lars}S. Larsson, Phys. Rev. \textbf{169}, 49 (1968).

\bibitem{LLQ}L.D. Landau and E.M. Lifshitz, {\it Quantum Mechanics: Non-Relativistic Theory}, (3rd. ed. Pergamon Press, New York (1976)), Chpt. VI.

\bibitem{Dirac}P.A.M. Dirac, \textit{The Principles of Quantum Mechanics} (4th ed., Oxford at the Clarendon Press, Oxford (UK) (1958)).

\bibitem{Liatom}R.G. Schlecht and D.W. McColm, Phys. Rev. \textbf{142}, 11 (1966).

\bibitem{CRC}\textit{CRC Handbook of Chemistry and Physics}, 95th Edition, Ed. W.M. Haynes, (Taylor and Francis Group, Boca Raton, FL, (2014-2015)).

\bibitem{FroWa2010}A.M. Frolov and David M. Wardlaw,  JETP {\bf 138}, 5 - 15 (2010).

\bibitem{KT}N.N. Kolesnikov and V.I. Tarasov, Yad. Phys. \textbf{35}, 609 (1982) [Sov. Phys. Nucl. Phys. \textbf{35}, 354 (1982)].  

\bibitem{PhysRevA1}P. J\"{o}nson and C. Froese Fischer, Phys. Rev. A \textbf{48}, 4113 (1993).

\bibitem{PRA2}S.N. Ray, T. Lee, and T.P. Das, Phys. Rev. A, \textbf{7}, 1469 (1973).

\bibitem{Chen1}C. Chen, J. At., Mol. and Opt. Phys., article ID 569876, 6 pp (2012).

\bibitem{IQC}D.R. Beck and C.A. Nicolaides, Int. J. Quant. Chem. \textbf{26}, suppl. 18, 467 (1984).

\bibitem{CPL}D. Sundholm and J. Olsen, Chem. Phys. Lett. \textbf{177}, 91 (1991).

\end{thebibliography}
\end{document}